\documentclass[a4paper,10pt, twocolumn]{article}
\usepackage{graphicx}

\title{\vspace{-4cm}\bf Spatial variations of the Sr I 4607~\AA~ scattering polarization signals at 
subgranular scale observed with ZIMPOL at GREGOR telescope}

\author{Sajal Kumar Dhara$^1$, Emilia Capozzi$^1$, Daniel Gisler$\textsuperscript{1,2}$ , Michele Bianda$^1$, Renzo Ramelli$^1$, \\
Svetlana Berdyugina$^2$, Ernest Alsina$^1$, and Luca Belluzzi$^1$}

\date{%
    $^1$Istituto Ricerche Solari Locarno (IRSOL), 6605 Locarno-Monti, Switzerland \\
    $^2$Kiepenheuer-Institut f\"ur Sonnenphysik (KIS), Sch\"oneckstrasse 6, Freiburg, Germany}
\begin{document}
\maketitle

\begin{abstract}
Sr I 4607~\AA~ spectral line shows one of the strongest scattering polarization signals in the visible solar
spectrum. The amplitudes of these signals are expected to vary at granular spatial scales.
This variation can be due to changes in the magnetic field intensity and orientation (Hanle effect) 
as well as due to spatial and temporal variations in the plasma properties. Measuring the spatial variation 
of such polarization signal would allow us to study the properties of the magnetic fields at subgranular 
region. But, the observations are challenging since both high spatial resolution and high 
spectropolarimetric sensitivity are required at the same time. To the aim of measuring these 
spatial variations at granular scale, we carried out a spectro-polarimetric measurement with 
the Zurich IMaging POLarimeter (ZIMPOL), at the GREGOR solar telescope at different limb distances on solar disk. 
Our results show a spatial variation of scattering linear polarization signals in Sr I 4607~\AA~ line at 
the granular scale at every $\mu$, starting from 0.2 to 0.8. The correlation between the polarization signal amplitude and 
the continuum intensity imply statistically that the scattering polarization is higher at the granular regions 
than in the intergranular lanes.
\end{abstract}

\section{Introduction}\label{sec:sec1}

Trujillo Bueno \& Shchukina (2007) foresee
spatial amplitude variations of the scattering polarization signals
in the Sr I 4607~\AA~line by solving the 3D radiative transfer problem of scattering 
line polarization in a realistic hydrodynamical model
 provided that granulation is resolved.
The origin of the effect could be related to variations in the magnetic field present in the granulation or the intergranular lanes, as well as 
local variations in the
anisotropy of the radiation field.
Spectro-polarimetric observations at $\mu$=0.3 ($\mu$ = cos $\theta$, where $\theta$ is the heliocentric angle) 
in the solar disk using ZIMPOL (Ramelli et al. 2014) at the
GREGOR telescope (Schmidt st al. 2012)
by Bianda et al.(2018) reported this spatial variation of the scattering polarization
in the Sr I 4607~\AA~line exits and the polarization signal is higher at the center of granules than in the intergranular lanes.
Recent magneto-convection simulations by del Pino Alem{\'a}n et al. (2018) shows that 
the theoretical scattering polarization signals are anticorrelated with the continuum intensity at all on-disk positions. 
Stokes filtergraph observations with Fast Solar Polarimeter (FSP) by Zeuner et al.(2018) at $\mu$ = 0.6 in the Sr I 4607.3~\AA~
line also support an anti-correlation between
the line core Stokes Q/I signals and the continuum intensity.

As a continuation of this work, we carried out another observing campaign during June, 2018 at GREGOR to measure these variations
at different limb distances on the solar disk.
We obtained several measurements at Sr I  4607~\AA~line at different positions on the solar disk. 
Seeing conditions allow us to
 achieve the required spatial resolution only few hours during observing days. We choose to observe quiet 
regions of the Sun at different limb distances.

In the next sections, we describe the observations at the GREGOR telescope and the data
reduction process. We study the possible presence of correlations between the spatial variations of the polarization signals
at different limb distances and those of the continuum intensity, which we have used as an
indicator of the locations of granular and intergranular lanes. At last we conclude our results.

\begin{figure*}[ht]
\centering
{\vspace*{-0.1\textwidth}
\includegraphics[width=0.35\textwidth,clip=]{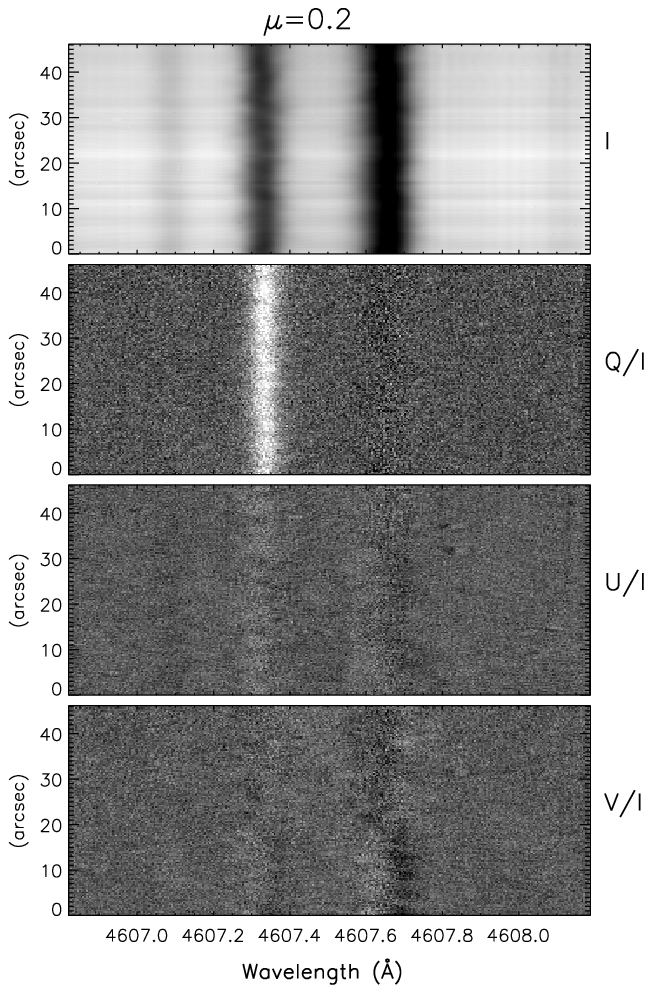}
   \hspace*{-0.04\textwidth}
 \includegraphics[width=0.35\textwidth,clip=]{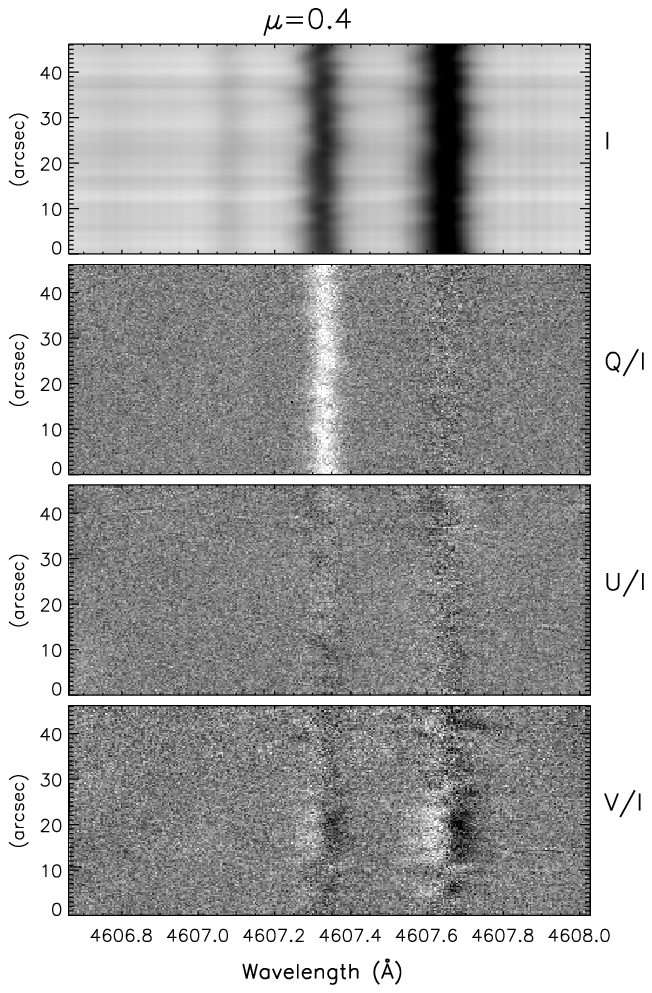}
  \hspace*{-0.04\textwidth}
 \includegraphics[width=0.35\textwidth,clip=]{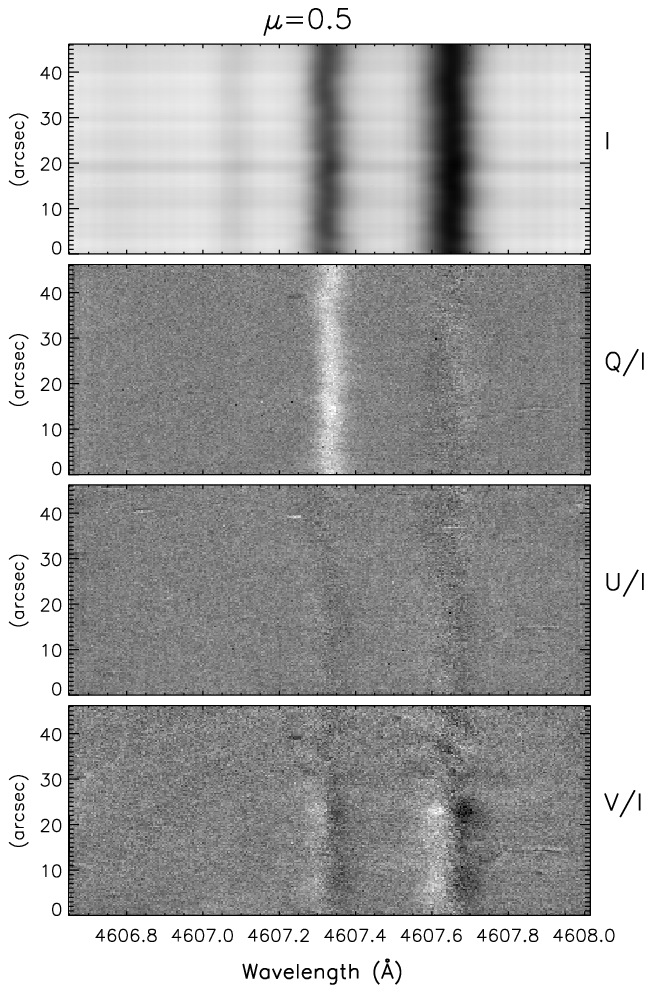}}
 \vspace{-0.05\textwidth}
\caption{Stokes images of a spectral interval around the Sr I 4607~\AA~line. The spatial direction spans $\sim$46'' on the solar disk. 
The observed regions were near the West limb at $\mu$ = 0.2, 0.4, and 0.5. The slit was placed parallel to the nearest limb. 
The reference direction for positive Stokes Q is the tangent to the
nearest solar limb. These images are the result of  $\sim$1.3, $\sim$1.7, and $\sim$3.3 minutes 
(close to the disk center, $\mu$=0.5 to 0.8, one need to average over longer time duration to improve the sufficient signal-to-noise
ratio.) observation average. 
The granulation pattern are visible in the intensity Stokes image,
in particular in the continuum. The Q/I image shows the scattering polarization peak in the core of the Sr I line.
Spatial variations at granular
scales of this peak can be observed. 
The typical
antisymmetric Zeeman patterns can be easily recognized in Stokes V/I for $\mu$ = 0.4 and 0.5.}
\label{fig:stokes_image1}
\end{figure*}

\section{Observations}\label{sec:sec2}

Our observation campaign at the GREGOR solar telescope was held between 13th June, 2018 to 27th June, 2018. 
The ZIMPOL system was installed at the GREGOR Spectrograph. 
The observations were performed at magnetically quiet regions of Sun, close to the West limb,
at different limb distances on the solar disk,
starting from $\mu$=0.2 to 0.8.
We have always used the adaptive optics system (Berkefeld et al. 2016) to get stable observations for required field-of-view positions. 
Close  to  the disk  center  ($\mu$ = 0.8  to  0.4),  the granular  structures  are  used  by  the  Shack 
Hartmann wavefront sensor of the adaptive optics system, 
while  near  to  limb  ($\mu$ = 0.3  to  0.2),  bright  plage  regions  are  used  to  lock  the 
regions over time during observations. 
The spectrograph slit was always placed in a magnetically quiet region, parallel to the nearest limb. The seeing quality 
fluctuated during our observations. The Fried parameter changes between $r_{0}$ = 3 cm
and $r_{0}$ = 10 cm. The spectrograph slit covers the 0.3'' (width of the slit) times 47''(length of the slit) on the solar area.
A ZIMPOL image has 140 pixels in spatial direction with a pixel resolution 0.33''. The estimated spatial resolution of the image for 
good observing conditions, is found
to be 0.6''. 
The spectral resolution of our observation is of $\sim$10 m\AA.
We used the GREGOR polarimetric calibration unit (GPU; Hofmann et al. 2012) that is placed at the the second focal point (F2).

\section{Results}\label{sec:sec3}

We follow the data reduction process, as mentioned in Bianda et al.(2018). %\cite{Bianda2018}. 
Figure~\ref{fig:stokes_image1} shows a sample of images 
obtained by averaging several frames 
 that were sequentially registered. The observed regions, shown in the Figure~\ref{fig:stokes_image1} 
were near the West limb at $\mu$ = 0.2, 0.4 and 0.5.
In the Stokes I images one can
recognize intensity variations along the spatial direction due to the granulation. The Q/I image shows the scattering polarization
peak in the Sr I 4607~\AA~spectral line. The polarization peak shows clear
spatial variations; detection of these variations was in fact the main goal of our
observations.

\begin{figure}[t]
\centering
{
\includegraphics[width=0.245\textwidth,clip=]{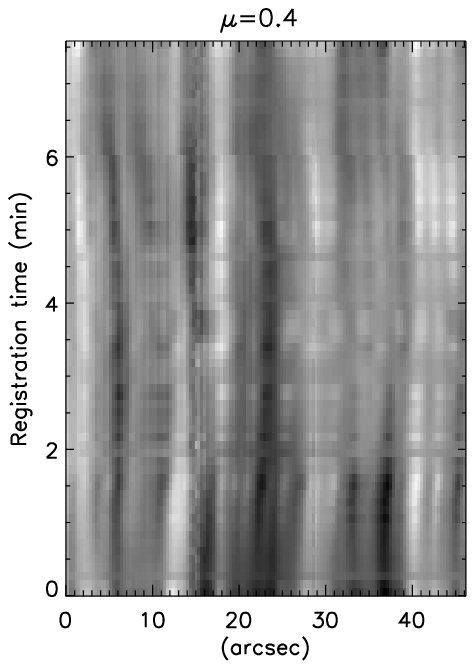}
 \hspace*{-0.04\textwidth}
\includegraphics[width=0.26\textwidth,clip=]{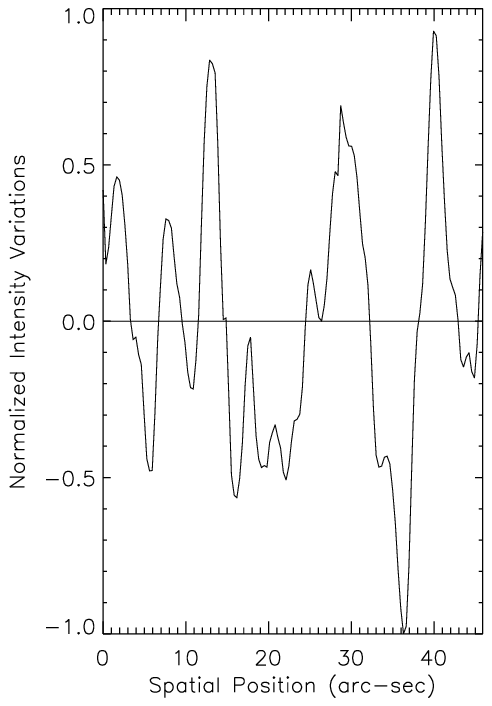}}
 \vspace{0.005\textwidth}
\caption{Left: Time space map shows the continuum profile along horizontal axis obtained from 
each Stokes I images (of total 68 frames) along the spectrograph slit. Each profile was obtained at
6.7 seconds, therefore, the temporal evolution of the field of view of the spectrograph slit is 7.6 minutes.
Right: Obtained normalized continuum intensity profile, averaged over 1.7 min from beginning, and then subtracted from median value.
This profile is used as a parameter to indicate the grannular (+ve) and inter-grannular (-ve) regions, in Figure~\ref{fig:scatter_plot}.
}
\label{fig:tme_space}
\end{figure}

At every observing position, a series of images are obtained. In order to improve the signal-to-noise ratio in the Stokes images, 
few subsequent frames are averaged over time.
At $\mu$=0.4, the data series contains 68 images (total observing time duration $\sim$ 7.6 min). %As previously mentioned, 
Taking into account the life time of granulation, we averaged 15 subsequent images of the time series, (integration time over about 1.7 minutes) to have a sufficient
 signal-to-noise ratio. 
 To evaluate whether this 
temporal averaging introduced significant spatial degradation in our
data, we generated a time space map of images seen by the spectrograph slit.
This map is shown in Figure~\ref{fig:tme_space} (left panel), where the continuum
intensity is reported in gray scale as a function of the spatial position along the slit (x-axis) and time along y-axis.
To locate the regions of granular and inter grannular lanes, a single continuum intensity profile is shown as an example in the
right panel of the Figure~\ref{fig:tme_space}. This profile is obtained by averaging over 1.7 minutes from the beginning, 
then subtracted from its median
value, and finally normalized with respect to the absolute maximum value. The positive and negative values represents the locations of
 the regions of granular and inter grannular lanes.
Data is now calculated from images
obtained by averaging the time series of 68 Stokes images with
a smoothing window width of 15 images, shifted in increments
of 15 images (to avoid data overlap). Few images are discarded due to poor seeing conditions during our observation.

\begin{figure*}[h]
\centering
{\vspace*{-0.15\textwidth}
\includegraphics[width=0.4\textwidth,clip=]{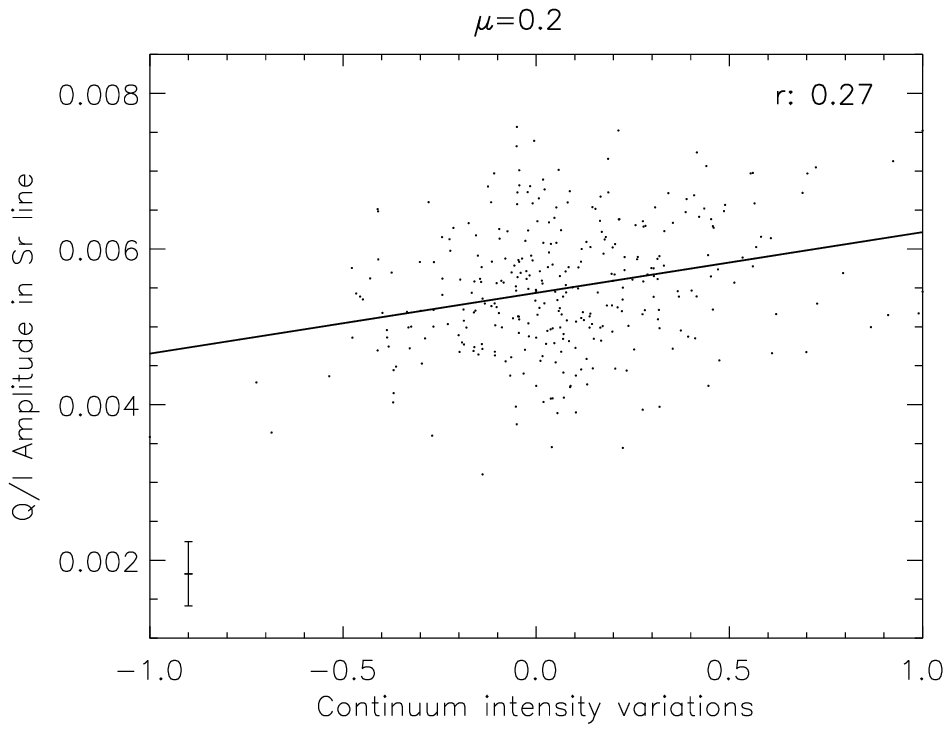}
   \hspace*{-0.02\textwidth}
 \includegraphics[width=0.4\textwidth,clip=]{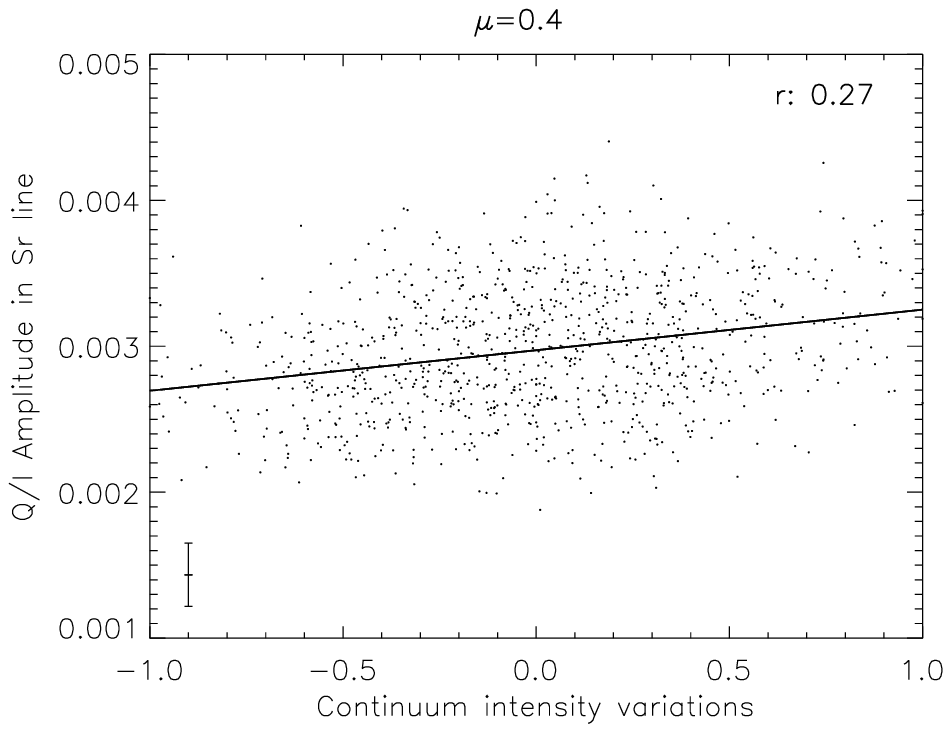}}
  %\hspace*{-0.04\textwidth}
 {\includegraphics[width=0.4\textwidth,clip=]{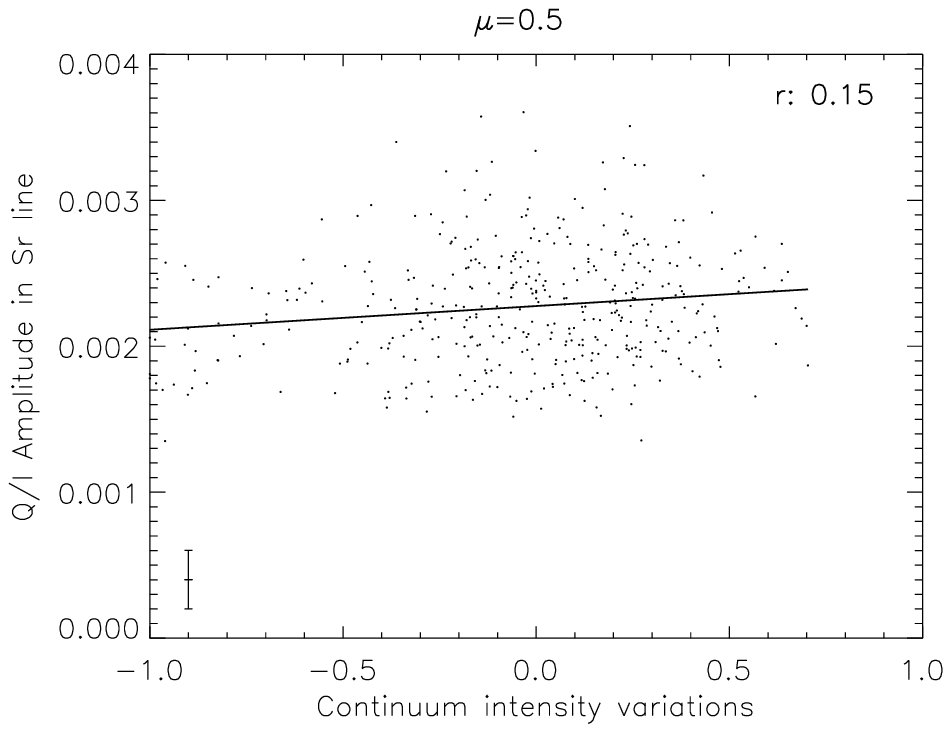}
\hspace*{-0.02\textwidth}
 \includegraphics[width=0.4\textwidth,clip=]{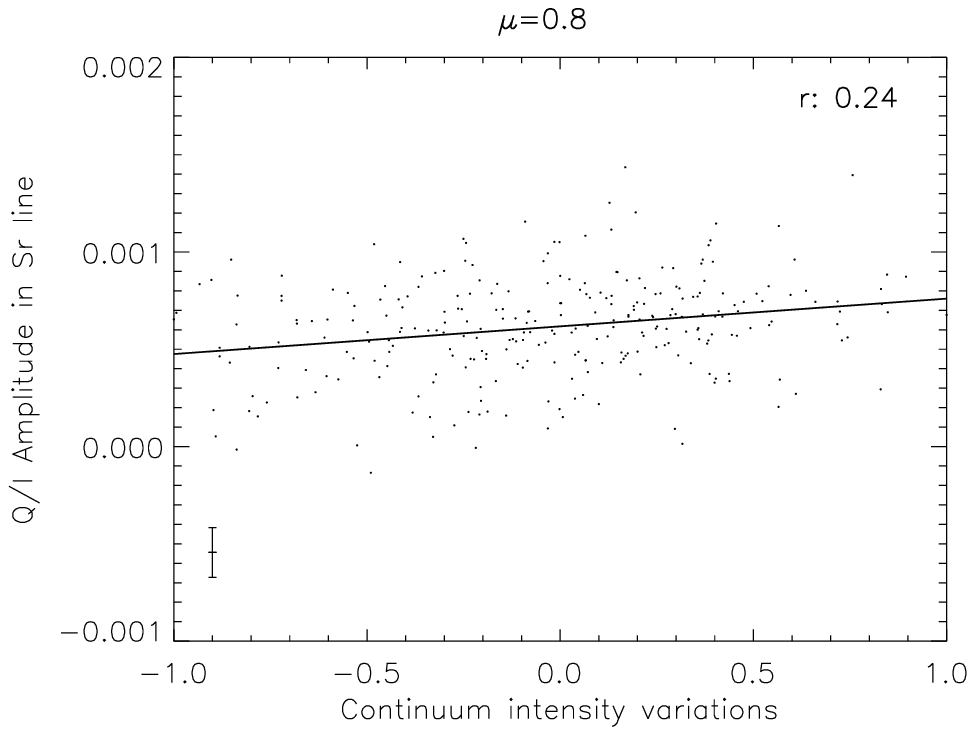}}
 \vspace{0.005\textwidth}
\caption{Scatter plot between the amplitudes of the Sr I 4607~\AA~Q/I peak and a continuum intensity 
indicatiung granulum (+ve value) or intergranulum (-ve value) regions at $\mu$= 0.2, 0.4, 0.5 and 0.8. 
The solid line shows linear regression that indicates larger polarization in the granulation. The Pearson correlation coefficient (PC) 
value is mentioned
in the graph. The estimated error for each points are shown in bottom left of each plot.}
\label{fig:scatter_plot}
\end{figure*}

The amplitude of each
Q/I signal peak of all the 15 averaged Stokes images (600 x 140 pixel) in the Sr I line at every spatial 
positions is calculated using a Gaussian fit to each of the 140 profiles. 
The scatter plot in Figure~\ref{fig:scatter_plot} shows how the Q/I peak signals
correlate with the continuum intensity at $\mu$ = 0.2, 0.4, 0.5 and 0.8. The solid line is a 
linear regression 
suggests a trend of increasing linear polarization in the granules compared to the intergranular regions.
The Pearson correlation coefficient (PC) of the individual plot is mentioned. We also obtained similar correlation results  
for other observing positions
($\mu$ = 0.6 and 0.7) as well, during good observing conditions. More details about the data reduction procedure
and the obtained results will be published in a refereed journal ({\it in preparation}) in future. %Dhara et al. ({\it in preparation}).
At $\mu$ = 0.45, 0.6 and 0.7, the observations were taken
during poor seeing conditions, where the correlation between the scattering 
polarization Q/I peak amplitudes and the continuum intensity disappeared.

\section{Conclusion}\label{sec:sec4}
We find spatial variations of
the Q/I scattering polarization signal of the Sr I 4607~\AA~ line,
measured at different limb distances (starting from $\mu$= 0.2 to 0.8, the direction of 
positive Stokes Q is parallel to the nearest solar limb)
on the solar disk. 
The obtained spatial scale of the variations
is comparable to the subgranular scale. There is a small correlation between the scattering polarization peak 
amplitudes of the Sr I 4607~\AA~line and the continuum
intensity. This implies, statistically that the polarization inside granular region is higher than in 
the intergranular lanes. 
Simulation by del Pino Alem{\'a}n et al. (2018) foresees an anti-correlation, but 
is based on high spatial resolution. Numerically deteriorating the signal-to-noise ratio and the spectral and spatial
resolutions of the simulated observations 
reproduces our results. Their theoretical study also suggests that, for 
investigating the scattering polarization signals of the Sr I 4607~\AA~line, a better instrument would be a
2D spectropolarimeter with a spectral resolution not worse than 20 m\AA, a spatial resolution $\sim$0.1'' and a polarimetric
sensitivity better than 10$^-$$^4$.

The observed results presented here, are mostly limited by 
seeing conditions, and also by the available photon statistics. 
A significant improvement of this study can be done by using imaging polarimetric
measurement at line core of Sr I line with high spatial resolution in a large telescope such as the DKIST.
In future, we are planning to develop a Fabry-Perot filter based polarimeter system,
for synoptic measurement at Sr I 4607~\AA~line, aiming at investigating the subgranular magnetic field, which could be a second generation 
instrument to be installed at DKIST.

\textit{\bf Acknowledgement}
\small
IRSOL is supported by the Swiss Confederation (SEFRI),
Canton Ticino, the city of Locarno and the local municipalities.
This research work was financed by SNF 200020\_169418. The 1.5-meter GREGOR solar telescope was built by a German 
consortium under the leadership of the Kiepenheuer-Institut fur Sonnenphysik 
in Freiburg with the Leibniz-Institut f\"ur Astrophysik Potsdam, the Institut f\"ur Astrophysik G\"ottingen, 
and the Max-Planck-Institut f\"ur Sonnensystem forschung in G\"ottingen as partners, and with contributions by the Instituto de
Astrofsica de Canarias and the Astronomical Institute of the Academy of Sciences of the Czech Republic.

\end{document}